\documentclass[twocolumn]{aastex61} 
\usepackage{amsmath}

\newcommand{\kms}{\ifmmode {\rm km\ s}^{-1} \else km s$^{-1}$\ \fi}
\newcommand{\ergs}{\ifmmode {\rm erg\ s}^{-1} \else erg s$^{-1}$\ \fi}
\newcommand{\lb}{\ifmmode L_{\rm Bol} \else $L_{\rm Bol}$\ \fi}
\newcommand{\ledd}{\ifmmode L_{\rm Edd} \else $L_{\rm Edd}$\ \fi}
\newcommand{\lx}{\ifmmode L_{\rm 2-10keV} \else  $L_{\rm 2-10keV}$\ \fi}
\newcommand{\ha}{\hbox{H$\alpha$}}
\newcommand{\hb}{\hbox{H$\beta$}}

\newcommand{\hg}{\hbox{H$\gamma$}}

\newcommand{\mbh}{\ifmmode M_{\rm BH}  \else $M_{\rm BH}$\ \fi}
\newcommand{\lv}{\ifmmode \lambda L_{\lambda}(5100\AA) \else $\lambda L_{\lambda}(5100\AA)$\ \fi}
\newcommand{\lbol}{\ifmmode L_{\rm Bol} \else $L_{\rm Bol}$\ \fi}

\newcommand{\te}{\,T_{\rm e} }

\newcommand{\oii}{\hbox{[O\,{\sc ii}]}}
\newcommand{\nii}{\hbox{[N\,{\sc ii}]}}
\newcommand{\sii}{\hbox{[S\,{\sc ii}]}}
\newcommand{\oiii}{\hbox{[O\,{\sc iii}]}}
\newcommand{\heii}{\hbox{[He\,{\sc ii}]}}
\newcommand{\nev}{[\rm Ne {\sc v}]}

\newcommand{\hii}{\hbox{H\,{\sc ii}}}

\newcommand{\oh}{\ifmmode 12+ \log({\rm O/H}) \else 12+log(O/H) \fi}
\newcommand{\mdot}{\ifmmode \dot{m} \else \dot{m} \fi }
\newcommand{\llog}{\ifmmode {\rm log} \else {\rm log} \fi }
\newcommand\simgt{\lower.5ex\hbox{\gtsima}}

\shorttitle{Metal-poor star-forming galaxies at $0.6<z<0.9$ in eBOSS }
\shortauthors{Gao, Bao, \& Yuan et al.}

\begin{document}

\title{Mass -- Metallicity Relation and Fundamental Metallicity Relation Of Metal-poor Star-forming Galaxies at $0.6<z<0.9$ from the {\small e}BOSS survey}


\author{YuLong Gao}
\affiliation{CAS Key Laboratory for Research in Galaxies and Cosmology, Department of Astronomy, University of Science and Technology of China, Hefei 230026, China}
\affiliation{School of Astronomy and Space Science, University of Science and Technology of China, Hefei 230026, China}

\author{Min Bao}
\thanks{Yulong Gao and Min Bao contributed equally to this work.}
\affiliation{Department of Physics and Institute of Theoretical Physics,
		  Nanjing Normal University, Nanjing 210023, China; yuanqirong@njnu.edu.cn}

\author{QiRong Yuan}
\affil{Department of Physics and Institute of Theoretical Physics,
		  Nanjing Normal University, Nanjing 210023, China; yuanqirong@njnu.edu.cn}

\author{Xu Kong}
\affiliation{CAS Key Laboratory for Research in Galaxies and Cosmology, Department of Astronomy, University of Science and Technology of China, Hefei 230026, China}
\affiliation{School of Astronomy and Space Science, University of Science and Technology of China, Hefei 230026, China}

\author{Hu Zou}
\affiliation{Key Laboratory of Optical Astronomy, National Astronomical Observatories, Chinese Academy of Sciences, Beijing 100012, China}

\author{Xu Zhou}
\affiliation{Key Laboratory of Optical Astronomy, National Astronomical Observatories, Chinese Academy of Sciences, Beijing 100012, China}

\author{Yizhou Gu}
\affil{Department of Physics and Institute of Theoretical Physics,
		  Nanjing Normal University, Nanjing 210023, China; yuanqirong@njnu.edu.cn}

\author{Zesen Lin}
\affiliation{CAS Key Laboratory for Research in Galaxies and Cosmology, Department of Astronomy, University of Science and Technology of China, Hefei 230026, China}
\affiliation{School of Astronomy and Space Science, University of Science and Technology of China, Hefei 230026, China}

\author{Zhixiong Liang}
\affiliation{CAS Key Laboratory for Research in Galaxies and Cosmology, Department of Astronomy, University of Science and Technology of China, Hefei 230026, China}
\affiliation{School of Astronomy and Space Science, University of Science and Technology of China, Hefei 230026, China}

\author{Chi Huang}
\affiliation{CAS Key Laboratory for Research in Galaxies and Cosmology, Department of Astronomy, University of Science and Technology of China, Hefei 230026, China}
\affiliation{School of Astronomy and Space Science, University of Science and Technology of China, Hefei 230026, China}

\begin{abstract}

The stellar mass-metallicity relation ($M_\star \-- Z$, MZR) indicates that the metallicities of galaxies increase with increasing stellar masses. The fundamental metallicity relation (FMR) suggests that the galaxies with higher star formation rates (SFRs) tend to have lower metallicities for a given stellar mass. To examine whether the MZR and FMR still hold at poorer metallicities and higher redshifts, we compile a sample of 35 star-forming galaxies (SFGs) at $0.6<z<0.9$ using the public spectral database ($\rm v5\_10\_0$) of emission-line galaxies from the extended Baryon Oscillation Spectroscopic Survey (eBOSS). These galaxies are identified for their significant auroral $\oiii\lambda4363$ emission line ($\rm S/N \geq 3$).  With the electronic temperature metallicity calibration, we find nine SFGs are extremely metal-poor galaxies with $\oh \leq 7.69 \ (1/10 \ Z_\odot)$. The metallicity of the most metal-deficient galaxy is $7.35\pm 0.09$ (about 1/20 $Z_{\odot}$). Compared with the SFGs with normal metallicities in local and high redshift universe, our metal-poor SFGs have more than ten times higher SFRs at a fixed stellar mass. We create a new mass -- SFR relation for these metal-poor galaxies at $0.6<z<0.9$. Due to the higher SFRs and younger stellar ages, our metal-poor SFGs deviate from the MZR and FMR in the local universe toward lower metallicities, confirming the existence of FMR, as well as the cosmic evolution of MZR and FMR with redshift. 

\end{abstract}

\keywords{galaxies: abundances --- galaxies:
distance and redshifts --- galaxies: evolution --- galaxies: ISM --- galaxies: photometry --- galaxies: starburst}

\section{Introduction}
\label{sec:intro}

The gas-phase heavy-element abundance in the interstellar medium (ISM), known as ``metallicity'',  is a fundamental quantity that reflects the evolutionary stage of galaxies.  The metallicity of a galaxy has been influenced by a lot of key processes in galaxy formation that remains to be deeply understood. The metal content within a galaxy seems to be primarily set by metal enhancement by star formation, but diluted in the short-term by the cosmological gas inflow and ejected by large-scale outflow via galactic winds \citep[e.g.,][]{Dalcanton2007,Dave2011,Lilly2013}. The inflowing gas provides the raw fuel for star formation on a longer timescale, while the galactic outflow enriches the ISM and inflowing gas. The baryons are cycling in and out of galaxies and may lead to a direct impact on the stellar masses ($M_\star$), metallicities ($Z$), and star formation rates (SFRs) of the galaxies. For this reason, the stellar mass-metallicity relation (MZR), the ``main sequence'' relation (MSR) between galaxy SFRs and stellar masses, and the fundamental metallicity relation (FMR, $M_\star$ -- SFR -- $Z$) may serve as observational constraints on models of galaxy evolution, which provides a better understanding of the build-up of galaxies across cosmic time.

The MZR, established by \cite{Lequeux1979a} and developed by a series of studies \citep[e.g.,][]{Tremonti2004, Mannucci2010, Andrews2013}, indicates a trend that the metallicities of galaxies increase with increasing stellar masses. \cite{Mannucci2010} and \cite{Andrews2013} found that metallicity has an anti­-correlation with SFR at a given stellar mass, and constructed the $M_\star$ -- SFR -- $Z$ relation (or FMR). However, \cite{Sanchez2013} and \cite{Barrera-Ballesteros2017} argued that MZR is independent on the SFR.   In addition, the evolution of MZR with redshift has also been explored for years \citep[e.g.,][]{Savaglio2005,Maiolino2008,Zahid2011,Yuan2013,Lian2015,Ly2016a}, since the larger telescopes and deeper spectroscopic surveys.  These studies suggested the existence of cosmic evolution for MZR, which means that galaxies at higher-$z$ universe tend to have lower metallicity at a fixed stellar mass.

There are a variety of methods to determine the metallicity \citep{Kewley2008}. Some calibrations are based on the photoionization models for $\hii$ regions by reproducing some emission line ratios, like $(\oii\lambda3727+\oiii\lambda\lambda4959,5007)/\hb$ \citep[R23,][]{Kobulnicky04}, $\nii\lambda6583/\oii\lambda3727$ \citep[N2O2,][]{Kewley2002}. Some other calibrations are empirical fits to the electronic temperature ($\te$) method with  strong-line ratios for $\hii$ regions and galaxies, like $(\oiii\lambda5007/\hb)/(\nii\lambda6583/\ha)$ and $\nii\lambda6583/\ha$ \citep[O3N2, N2,][]{Pettini2004}.  However, there are some problems when using these strong-line metallicity calibrations. For example, the MZRs with different calibrations have different shapes and normalization values \citep{Kewley2008}. For high-$z$ star-forming galaxies, these calibrations may not be valid, since their physical conditions (e.g., gas density, ionization, N/O abundance) of the interstellar gas are significantly different from those in local universe \citep{Ly2016a}.

The metallicity calibration with the electronic temperature of ionized gas, using the $\oiii\lambda4363/\oiii\lambda5007$ ratio, which is also called $\te$ method \citep{Aller1984}, is considered as the most reliable approach to determine the gas-phase oxygen-to-hydrogen (O/H) abundance \citep{Izotov2006}.  The weak $\oiii\lambda4363$, produced by $\hii$ regions with high enough temperature, can only be detected in metal-poor emission line galaxies. For the local universe, some previous studies have made enormous efforts to enlarge the sample size of metal-poor galaxies with $\te$--based metallicities \citep[e.g.,][]{Izotov2006,Berg2012,Izotov2012,Ylgao2017,Hsyu2018a}.  \cite{Izotov2006} found six new extremely metal-poor galaxies (XMPGs; $\oh \leq 7.69$) from 310 emission line galaxies (ELGs) with $\rm S/N(\oiii\lambda4363) > 2$ in Sloan Digital Sky Survey (SDSS) Data Release (DR) 3 dataset. \cite{Berg2012} also investigated 19 new metal-poor galaxies with $\rm S/N(\oiii\lambda4363) > 4$ using the MMT telescope. In the SDSS DR7, \cite{Izotov2012} found seven metal-deficient galaxies with $\oh \leq 7.35$. Based on the photometric colors and morphologies in SDSS DR12, \cite{Hsyu2018a} determined 45 blue compact dwarf galaxies (BCDs) with $\oh \leq 7.65$. However, the weak $\oiii\lambda4363$ is even harder to detect for galaxies at $z \ge 0.2$. Some studies used large telescopes, such as Keck II, MMT and VLT, to search the metal-poor galaxies with $\oiii\lambda4363$ detection at $0.5 \le z \le 1.0$ \citep[e.g.,][]{Kakazu2007,Amor2014,Ly2014,Ly2015a,Ly2016a}. In total, the number of the metal-poor galaxies with $\rm S/N(\oiii\lambda4363) \ge 3$ is less than 300.

The metal-poor galaxies play an essential role in understanding the galaxy evolution, especially for the galaxies at the early stage of evolution, or for those evolve slowly. The low metallicities of these galaxies also suggest that they might have significant metal-poor gas inflows or metal-enriched gas outflows.  The interstellar medium (ISM) in these galaxies is nearly pristine and could shed light on the ISM properties at the galaxy formation time. Furthermore, because these galaxies located in the very early stage of chemical evolution, a large sample of metal-poor galaxies is necessary to improve the determination of primordial $^4$He abundance \citep{Izotov2004}  predicted by the standard Big Bang nucleosynthesis model.

In this work, we try to search the new metal-poor galaxy candidates at $0.6<z<0.9$ by their $\oiii\lambda4363$ line from the eBOSS survey, and then explore the cosmic evolution of MZR and FMR with metallicities derived using direct method. In our next work, using the strong-line metallicity calibrations, Huang et al. (2018, in prep) will divide all the ELGs in eBOSS survey into different bin regions, based on stellar mass, SFR and the 4000$\rm \AA$ break ($D_{\rm n}4000$), to mainly explore the cosmic evolution of MZR and FMR.

This paper is organized as follows. In Section \ref{sec:data}, we describe the methodology for detecting and measuring nebular emission lines, the measurements of metallicity and stellar mass, the selection for star-forming galaxies and SFR determination. In Section \ref{sec:results}, we provide the results and discussion about the main sequence relation, the MZR and FMR based our metal-poor galaxies.  Finally, we present the main results in Section \ref{sec:summary}.  Throughout this paper, we adopt the solar metallicity ($Z_\odot$) as $\oh = 8.69$ \citep{Allende2001} and a flat $\Lambda$CDM cosmology with $\Omega_\Lambda=0.7$, $\Omega_{\rm m}=0.3$, and $H_0=70$ km s$^{-1}$ Mpc$^{-1}$.

\section{The data analysis}
\label{sec:data}

\subsection{The eBOSS Overview}

The Extended Baryon Oscillation Spectroscopic Survey \citep[eBOSS,][]{Dawson2016}, one of the three core programs in the SDSS-IV \citep{Blanton2017}, will map the distribution of galaxies and quasars at redshift $z \sim$ 0.6 -- 3.5 and improve constraints on the nature of dark energy. This survey started in 2014 July and will create the largest volume survey of the universe to date when finished until Spring 2020. eBOSS will target about 300,000 luminous red galaxies (LRGs) over 7500 deg$^2$ (0.6 $< z <$ 0.8), 189,000 ELGs over 1000 deg$^2$ (0.6 $< z <$ 1.0) and 73,000 quasars over 7500 deg$^2$ (0.9 $< z <$ 3.5), and then provide the spectra with coverage of 3600 -- 10300$\rm \AA$ at a resolution of  $R \sim$ 2000. The eBOSS team has delivered a public Value Added Catalogue \footnote{https://data.sdss.org/sas/dr14/eboss/spectro/redux/v5\_10\_0/} (VAC; `spAll-v5\_10\_0.fits') together with spectral data covering $\sim$ 2480 deg$^2$ footprint. These data include all sources observed in the first two years of SDSS­-IV operations \citep{Abolfathi2018}. From the 3,008,000 objects in the VAC, we select 345,584 galaxies as the parent sample based on  the following criteria: (1) $zwarning = 0$, (2) $class = `GALAXY'$, (3) $platequality \neq `bad'$ and (4) $0.6 < z < 1.0$.

\subsection{Spectral Fitting and Emission-line Measurements}

With the spectra of these 345,584 galaxies in hand, we first use the color excess E(B -- V) map of the Milky Way \citep{Schlegel1998} to correct the Galactic reddening for these spectra. Then we mask out the optical emission lines and reproduce the underlying stellar continuum using STARLIGHT code \citep{CidFernandes2005}, adopting the combination of 45 single stellar populations (SSPs) from \cite{Bruzual2003} (BC03) model and the \cite{Chabrier2003} initial mass function (IMF). These SSPs consist of 15 ages in the range of 1 Myr -- 13 Gyr and three different metallicities (0.01, 0.02, 0.05).  After subtracting the stellar continuum from the spectrum, we use the multiple Gaussians profiles to measure the fluxes of emission lines (e.g., $\rm \oii\lambda3727$, $\hg$, $\oiii\lambda4363$, $\hb$, $\oiii\lambda\lambda4959,5007$) with IDL package MPFIT \citep{Markwardt2009}. We also estimate the signal-to-noise ratios (S/N) for these emission lines following the method in \cite{Ly2014}, which determines the flux with
\begin{equation}
  {\rm Flux} = \sum_{-2.5\sigma_G}^{+2.5\sigma_G} \left[f(\lambda-l_C)-\left<f\right>\right] \times l\arcmin,
\end{equation}
and assumes the noise as
\begin{equation}
  {\rm Noise} = \sigma(f) \times l\arcmin \times \sqrt{N_{\rm pixel}}.
\end{equation}
Here, $\sigma_G$ is the Gaussian width, $\left<f\right>$ and $\sigma(f)$ are the median value and standard deviation of flux densities within 200$\rm \AA$-wide region of spectral continuum, excluded the skylines and emission lines, repectively. $l_C$ is the center wavelength of emission lines, $l\arcmin$ is the spectral dispersion, and $N_{\rm pixel}=5\sigma_G/l$\arcmin. Finally, we assume the ``Case B'' recombination model and the \cite{Calzetti2000} reddening formalism to derive the color excesses E(B -- V) and correct the fluxes for these emission lines. We set E(B -- V) to zero if the observed ratio $\hg/\hb$ is higher than the intrinsic ratio $(\hg/\hb)_0$ = 0.468 \citep{Hummer1987}.

\subsection{Primary Sample Selection}
\label{sub:sample}

In order to get accurate SFRs with emission line luminosities and reliable metallicities with electronic temperature, we select these spectra with the following criteria: S/N($\oiii\lambda4363$) $\ge 3$, S/N($\oii\lambda3727$) $\ge 5$, S/N($\oiii\lambda5007$) $\ge 10$ and S/N($\hb$) $\ge 5$. As a result, we get a subsample of 319 galaxies.

\subsection{The Stellar-mass Determination From SED Modeling}

Since the coverage of rest-frame spectra in our subsample is very narrow ($< 3000 \rm \AA$), the stellar masses derived by fitting stellar continuum with STARLIGHT code are not credible. We perform the spectral energy distribution (SED) fitting with multi-photometric ($g,r,z, W1(3.4 \mu m), W2(4.6\mu m)$ bands) measurements using the IDL code library FAST developed by \cite{Kriek2009}.

Because the magnitude values of $z$ band in SDSS have significant uncertainties, we preferentially collect the MODELFLUX values of $g,r,z, W1, W2$ bands for 158/319 galaxies from the Legacy Surveys \footnote{http://legacysurvey.org/} \citep{Lang2014,Raichoor2017,Zou2018}. The Legacy Surveys produce an inference model catalog of the sky from a series of optical and infrared imaging data. The MODELFLUX values of $g,r,z$ bands for the rest 161/319 galaxies are collected from the eBOSS VAC Catalogue, while the MODELMAG values of $W1, W2$ bands for 78/161 galaxies are derived by cross-matching with the All WISE Source Catalog \citep{Wright2010}. Besides, we correct the broadband ($g,r,z$ bands) photometry by subtracting the contribution of strong emission lines from our spectroscopic, which has been used in previous studies \citep[e.g.,][]{Atek2011, Pirzkal2013, Izotov2014, Ly2014}.

In the SED fitting, we adopt the \cite{Chabrier2003} IMF, the \cite{Bruzual2003} stellar templates with four metallicities (0.004, 0.008, 0.02, 0.05) and an exponentially decreasing star formation model $\rm SFR \varpropto e^{- t/\tau}$ to synthesize magnitudes. We set the stellar population ages ranging from 0 to 10 Gyr and also assume the \cite{Calzetti2000} reddening formalism allowing E(B -- V) to vary from 0.0 to 2.0. Varying the input photometry with their photometric errors, we repeat the SED fitting for 1000 times. The stellar mass and its error are assumed as the median value and one sigma value in the distribution of fitting results. The median and standard deviation of the errors on stellar mass estimation in SED fitting are 0.26$\pm$0.18 dex, while the uncertainties in SSP selections and mass-to-light ratio are about 0.28 dex (Section \ref{subsec:discussion_z_mass}, in detail) and  0.1 dex \citep{Bruzual2003}, respectively.

\subsection{$T_{\rm e}$-based Metallicity Determination}
\label{sub:metallicity}

To determine the gas-phase metallicity for our SFGs with significant detection of $\oiii\lambda4363$, we use the $T_{\rm e}$ method \citep[e.g.,][]{Aller1984,Izotov2012,Ly2016,Bian2018a}. In general, one needs to use the $\sii\lambda\lambda6717,6731$ doublet lines to determine the electron densities ($n_e$) and the electron temperatures $T_{\rm e}$. However, since the $\sii$ doublet lines are redshifted out of our spectral coverage, we assume the $n_e$ as 100 $\rm cm^{-3}$ like used in \cite{Ly2014}. Same as our previous work \citep{Ylgao2017,Lin2017}, we use the python package PYNEB \footnote{http://www.iac.es/proyecto/PyNeb/} developed by \cite{Luridiana2015} to calculate the metallicity. In the calculation process, we set the atomic recombination data and atomic collision strength data for a series of ions (O$^+$, O$^{++}$) \citep[][in detail]{Ylgao2017}. We also follow an iterative method in \cite{Nicholls2014} to determine the temperature of O$^+$ region $T_{\rm e}$($\oii$) with the temperature of O$^{++}$ region $T_{\rm e}$($\oiii$) and the total oxygen abundance, $\oh$.

For the uncertainty in metallicity determination, we simulate the emission lines according to the measured fluxes and errors, and repeat the metallicity calculation for 1000 times. We regard  the median value of the 1000 measurements as the measurement value, and the one sigma value of the distribution as the error. The median and standard deviation of uncertainties for metallicity are 0.15$\pm$0.07 dex.

\begin{figure}
\epsscale{1.0}
\plotone{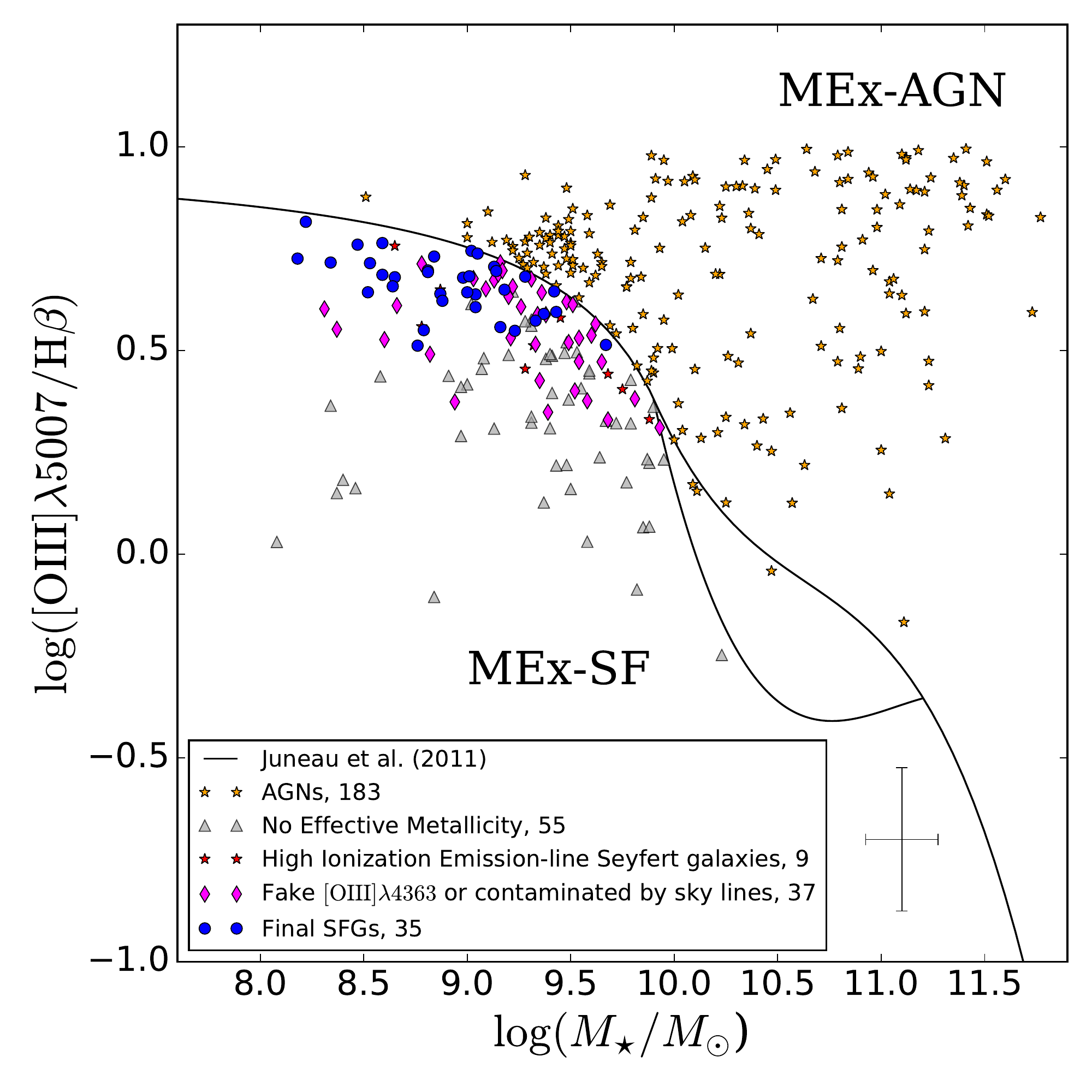}
\caption{
Flux ratio $\oiii/\hb$ as a function of stellar mass (i.e., the MEx diagram). 183 (green stars) and nine (red stars) galaxies are identified as AGNs via the black solid \cite{Juneau2011} demarcation lines and the presence of high ionization lines $\nev\lambda3425$ and $\heii\lambda4686$, respectively.  37 galaxies (magenta diamonds) are found with fake $\oiii\lambda4363$ detection or severely contaminated by OH skylines.  Blue filled circles denote the SFG sample. 55 galaxies (gray triangles) have ineffective metallicities. Blue filled circles denote our final SFG sample. The error bar shows the median values of measurement uncertainties for stellar mass and $\oiii/\hb$ ratio.
\label{fig:MEx}}
\end{figure}

\subsection{Exclusion of AGNs and Final Sample Selection}
\label{sub:MEx}

Due to the absence of the emission lines $\ha$ and $\nii\lambda6583$ for all spectra in the subsample, we exclude the galaxies that harbor active galactic nucleus (AGNs) using the ``Mass-extinction'' diagram \citep[MEx;][]{Juneau2011,Trump2013}, instead of the BPT diagram \citep{Baldwin1981, Kewley2001, Kauffmann2003}.  There are 136/319 galaxies located in the star-forming region. However, taking the $T_{\rm e}$-based metallicity in Section \ref{sub:metallicity} into consideration, we have 81/136 galaxies with effective metallicities.  Then we inspect the spectra of rest 81 galaxies visually, and exclude 9/81 Seyfert galaxies with high ionization lines $\nev\lambda3425$ and $\heii\lambda4686$ \citep{Izotov2012, Ly2014}, as well as 37/81 spectra with fake $\oiii\lambda4363$ detection or contaminated by OH skylines. Finally, our final SFG sample consists of 35 galaxies.

\begin{figure*}
\epsscale{0.95}
\plotone{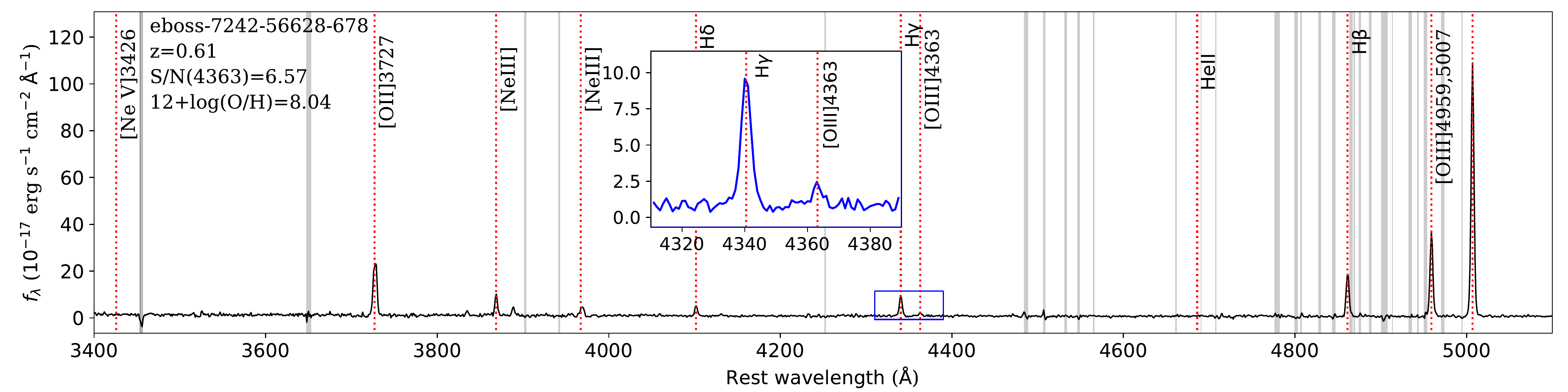}
\plotone{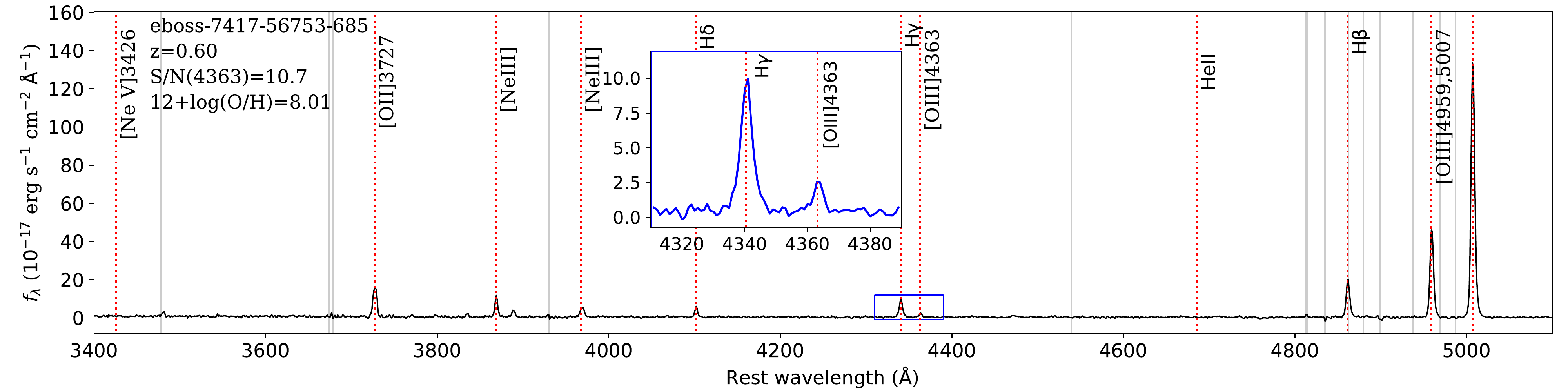}
\plotone{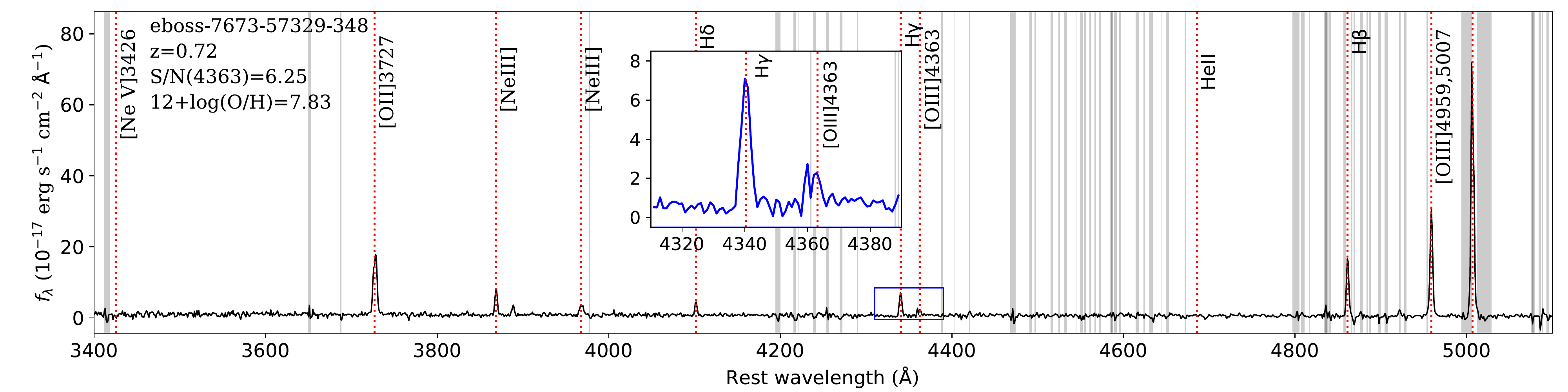}
\plotone{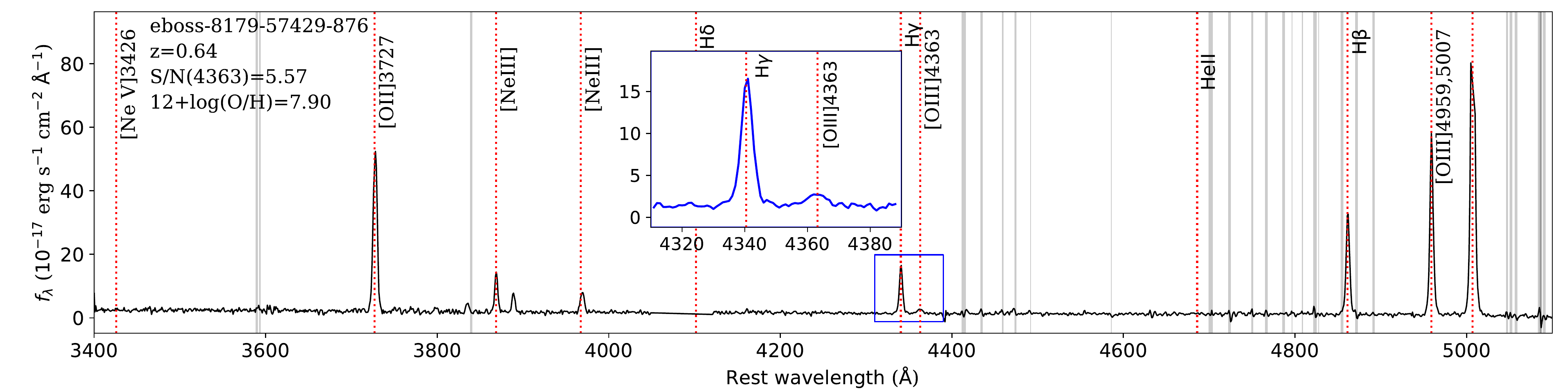}
\plotone{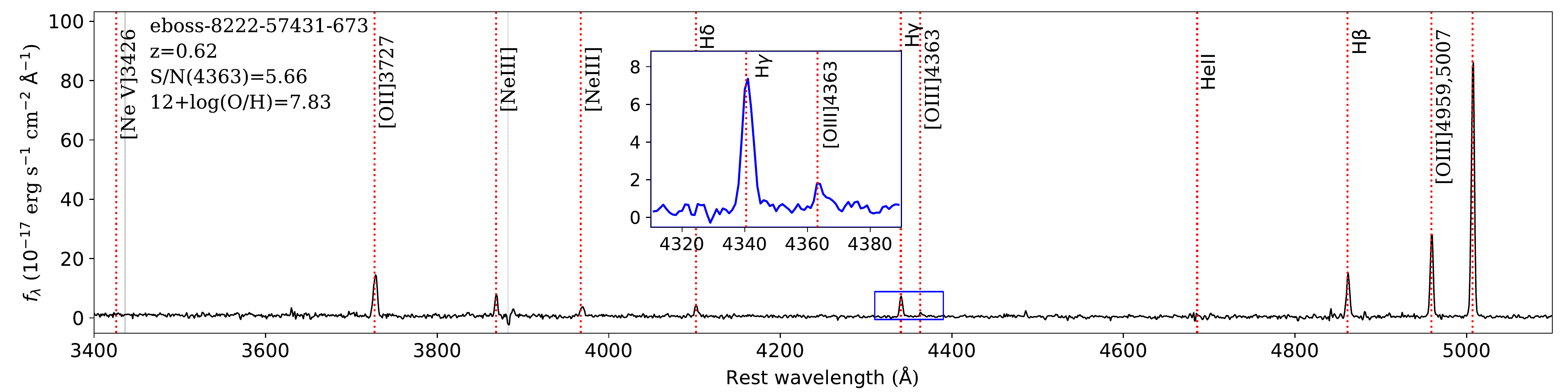}
\plotone{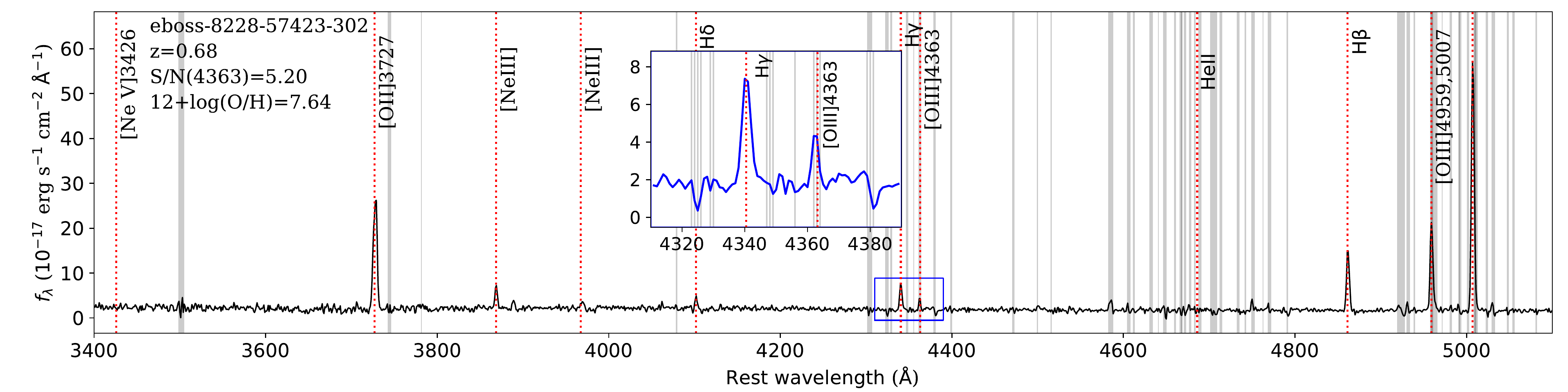}
\caption{ The eBOSS spectra for 6 representative $\oiii\lambda4363$-detected SFGs,
which span a rest-frame wavelength range from 3400$\rm \AA$ to 5100$\rm \AA$, cover the domains of $\nev\lambda3425$ to $\oiii\lambda5007$. The insets provide a zoomed view of $\hg$ and $\oiii\lambda4363$. The spectral ID ($plate\--mjd\--fiberid$), redshift, S/N($\oiii\lambda4363$) and $T_{\rm e}$-based metallicity are also given in the top left. The vertical gray shaded bands denote the OH skylines in the manner that the darker bands indicate the stronger skylines.
\label{fig:spec}}
\end{figure*}

The sample selection is shown in the Figure \ref{fig:MEx}. All symbols represent the 319 candidates in subsample selected with signal-to-noise of emission lines in Section \ref{sub:sample}.  Green and red stars represent the 183 AGNs in MEx-AGN region and nine Seyfert galaxies, respectively.  Gray triangles and magenta diamonds show the 55 galaxies without effective metallicities and 37 galaxies with fake $\oiii\lambda4363$ detection or contaminated by OH skylines. Blue filled circles denote our final SFG sample. Figure \ref{fig:spec} shows the eBOSS spectra for six representative $\oiii\lambda4363$-detected SFGs, which span a rest-frame wavelength range from 3400$\rm \AA$ to 5100$\rm \AA$. For each panel, the spectral ID ($plate\--mjd\--fiberid$), redshift, S/N($\oiii\lambda4363$) and $T_{\rm e}$-based metallicity are also given in the top left. The OH skylines are denoted by the vertical gray shaded bands in the manner, that the darker bands indicate the stronger skylines.

\subsection{SFR Estimation From $\hb$}
\label{sub:sfr}

In addition to stellar mass and oxygen abundances, we can estimate the dust-corrected SFRs using the $\hb$ luminosities $L_{\rm cor}(\hb)$. Assuming a \cite{Chabrier2003} IMF and solar metallicity, the SFR calculated in \cite{Kennicutt1998} can be written as
\begin{equation}
\label{eq:sfr_ha}
\begin{split}
{\rm SFR}(M_{\sun} {\rm yr^{-1}}) = 4.4 \times 10^{-42} \times (\ha/\hb)_0 \\
 \times L_{\rm cor}(\hb)(\ergs), 
\end{split}
\end{equation} \\ 
where the intrinsic ratio $(\ha/\hb)_0 = 2.86$. However, for the galaxies with lower metallicities, the SFRs estimated by the Eq.\ref{eq:sfr_ha} will be overestimated since the greater escape of ionizing photons from more metal-poor O star atmospheres. \cite{Ly2016} gave the metallicity-corrected SFRs as
\begin{equation}
\label{eq:sfr_cor}
{\rm log(SFR_{cor})} = {\rm log(SFR)} + 0.39 {\rm y} + 0.127{\rm y^2},	
\end{equation}
where $\rm y = log(O/H) + 3.31$.

We summarize the basic information, properties and measurements of the final $\oiii\lambda4363$-detected SFGs selected from eBOSS in Table \ref{table1}. We find nine SFGs are extremely metal-poor with $\oh \leq 7.69$ (i.e., poorer than 1/10 $Z_\sun$). The metallicity for the most metal-deficient galaxy (ID: 7396-56809-134, $z = 0.68$) is $7.35\pm 0.09$, about 1/20 $Z_{\sun}$.

\tabcolsep 0.035in
\begin{longrotatetable}
\linespread{0.8}
\begin{deluxetable*}{crrccccrrcccccc}
\tablecaption{The properties and measurements of the $\oiii\lambda4363$-detected SFGs from eBOSS \label{table1}}
\tablewidth{700pt}
\tabletypesize{\scriptsize}  
\tablehead{
\colhead{ID} & \colhead{R.A.} &
\colhead{DEC.} & \colhead{$z$} &
\colhead{$E$(B-V)} & \colhead{S/N} &
\colhead{$D_{\rm n}4000$} & \colhead{EW($\hb$) } &
\colhead{$\hb$} &
\colhead{$\frac{\oiii\lambda5007}{\hb}$} &
\colhead{$\frac{\oiii\lambda5007}{\oiii\lambda4363}$} &
\colhead{log($M_{\star}/M_{\odot}$)} &
\colhead{log($\frac{{\rm SFR}[\hb]}{M_{\odot} {\rm yr}^{-1}}$)} &
\colhead{$\log(T_{\rm e}/{\rm K})$ } &
\colhead{$12+\log({\rm O/H})$} \\
\colhead{(plate-mjd-fiberid)}  & \colhead{(deg)} & \colhead{(deg)}   & \colhead{}   & \colhead{(mag)} & \colhead{($\lambda4363$)} & \colhead{} & \colhead{(\AA)}   &  \colhead{($10^{-17}\ergs$)}  & \colhead{}   & \colhead{}   &  \colhead{}  &  \colhead{}  &  \colhead{}  & \colhead{} \\
\colhead{(1)} & \colhead{(2)}& \colhead{(3)}& \colhead{(4)}& \colhead{(5)}& \colhead{(6)}& \colhead{(7)}& \colhead{(8)}& \colhead{(9)}& \colhead{(10)}& \colhead{(11)} & \colhead{(12)} & \colhead{(13)} & \colhead{(14)} & \colhead{(15)}  \\
}
\startdata
7863-56975-800  &    3.0797  &   -1.0267  &    0.62  & $   0.40 \pm   0.15 $ &        3.74  &      0.90   &   44.36  & $     335.36 \pm  12.99  $ & $ 3.53 \pm     0.16     $ & $ 49.35 \pm 11.52 $ & $     9.23 \pm 0.05 $ & $   1.83 \pm   0.04     $ & $   4.18 \pm   0.12 $ & $   7.88 \pm   0.15 $  \\
7673-57329-348  &   11.1328  &   25.6580  &    0.73  & $   0.17 \pm   0.13 $ &        6.26  &      0.75   &  107.43  & $     106.90 \pm   5.00  $ & $ 5.38 \pm     0.38     $ & $ 42.53 \pm 18.58 $ & $     8.84 \pm 0.49 $ & $   1.50 \pm   0.05     $ & $   4.22 \pm   0.12 $ & $   7.84 \pm   0.13 $  \\
6288-56192-216  &   13.3805  &   23.2618  &    0.77  & $   0.00 \pm   0.15 $ &        4.69  &      0.79   &   69.74  & $      59.83 \pm   3.48  $ & $ 4.77 \pm     0.43     $ & $ 36.10 \pm 18.75 $ & $     8.98 \pm 0.47 $ & $   1.31 \pm   0.06     $ & $   4.25 \pm   0.15 $ & $   7.69 \pm   0.16 $  \\
7241-56606-699  &   22.6589  &   21.7292  &    0.68  & $   0.00 \pm   0.19 $ &        3.54  &      0.79   &   81.31  & $      27.05 \pm   0.85  $ & $ 5.08 \pm     0.24     $ & $ 48.94 \pm 20.22 $ & $     9.13 \pm 0.34 $ & $   0.83 \pm   0.03     $ & $   4.19 \pm   0.16 $ & $   7.81 \pm   0.20 $  \\
7247-56626-567  &   23.2868  &   19.3943  &    0.62  & $   0.23 \pm   0.11 $ &        3.52  &      0.73   &   86.44  & $     128.37 \pm   2.40  $ & $ 3.93 \pm     0.10     $ & $ 67.19 \pm 10.29 $ & $     9.43 \pm 0.50 $ & $   1.41 \pm   0.02     $ & $   4.13 \pm   0.08 $ & $   7.97 \pm   0.14 $  \\
7242-56628-678  &   23.6418  &   22.3382  &    0.61  & $   0.00 \pm   0.09 $ &        6.58  &      0.72   &  126.87  & $      76.68 \pm   1.28  $ & $ 5.20 \pm     0.12     $ & $ 70.55 \pm 9.46 $ & $     8.34 \pm 0.45 $ & $   1.18 \pm   0.02     $ & $   4.12 \pm   0.05 $ & $   8.05 \pm   0.07 $  \\
7248-56630-490  &   23.9039  &   18.4817  &    0.62  & $   0.44 \pm   0.16 $ &        3.60  &      0.91   &   25.10  & $     203.34 \pm   5.10  $ & $ 3.26 \pm     0.11     $ & $ 26.91 \pm 3.86 $ & $     9.67 \pm 0.08 $ & $   1.62 \pm   0.03     $ & $   4.33 \pm   0.12 $ & $   7.62 \pm   0.14 $  \\
5119-55836-950  &   30.8324  &   17.0539  &    0.65  & $   0.00 \pm   0.12 $ &        3.67  &      0.86   &   69.43  & $      81.03 \pm   1.79  $ & $ 3.61 \pm     0.12     $ & $ 47.84 \pm 10.43 $ & $     9.16 \pm 0.05 $ & $   1.26 \pm   0.02     $ & $   4.19 \pm   0.10 $ & $   7.81 \pm   0.13 $  \\
8736-57400-153  &   31.6903  &    3.9447  &    0.73  & $   0.00 \pm   0.15 $ &        4.43  &      0.81   &   62.09  & $      48.51 \pm   2.44  $ & $ 4.78 \pm     0.29     $ & $ 34.25 \pm 5.25 $ & $     8.65 \pm 0.06 $ & $   1.16 \pm   0.05     $ & $   4.27 \pm   0.07 $ & $   7.63 \pm   0.12 $  \\
4261-55503-832  &   40.4958  &    3.1851  &    0.82  & $   0.02 \pm   0.14 $ &        4.37  &      0.70   &  308.96  & $      63.42 \pm   1.71  $ & $ 4.80 \pm     0.18     $ & $ 36.03 \pm 5.20 $ & $     9.28 \pm 0.11 $ & $   1.41 \pm   0.03     $ & $   4.25 \pm   0.07 $ & $   7.61 \pm   0.08 $  \\
8858-57450-666  &  128.5331  &   38.5511  &    0.67  & $   0.27 \pm   0.17 $ &        5.69  &      0.79   &   83.27  & $     116.07 \pm   1.24  $ & $ 4.95 \pm     0.08     $ & $ 35.64 \pm 2.65 $ & $     9.14 \pm 0.13 $ & $   1.45 \pm   0.01     $ & $   4.26 \pm   0.11 $ & $   7.73 \pm   0.14 $  \\
8828-57445-880  &  130.9153  &   40.2271  &    0.73  & $   0.21 \pm   0.12 $ &        3.00  &      0.81   &   76.28  & $     191.35 \pm   4.31  $ & $ 4.34 \pm     0.13     $ & $ 49.30 \pm 6.89 $ & $     9.04 \pm 0.35 $ & $   1.76 \pm   0.02     $ & $   4.18 \pm   0.08 $ & $   7.86 \pm   0.13 $  \\
7315-56685-689  &  140.7128  &   46.8123  &    0.64  & $   0.12 \pm   0.12 $ &        4.52  &      0.80   &   77.82  & $      97.45 \pm   2.82  $ & $ 4.39 \pm     0.17     $ & $ 47.22 \pm 8.13 $ & $     8.52 \pm 0.11 $ & $   1.33 \pm   0.03     $ & $   4.19 \pm   0.07 $ & $   7.83 \pm   0.10 $  \\
7283-57063-843  &  146.3134  &   54.8753  &    0.62  & $   0.11 \pm   0.11 $ &        5.30  &      0.79   &   82.87  & $     107.48 \pm   1.98  $ & $ 4.55 \pm     0.12     $ & $ 45.56 \pm 5.86 $ & $     8.64 \pm 0.05 $ & $   1.34 \pm   0.02     $ & $   4.20 \pm   0.08 $ & $   7.82 \pm   0.12 $  \\
8823-57446-847  &  148.8682  &   39.5981  &    0.83  & $   0.39 \pm   0.12 $ &        4.44  &      0.87   &   71.88  & $     287.18 \pm   3.33  $ & $ 4.36 \pm     0.08     $ & $ 38.32 \pm 5.30 $ & $     8.87 \pm 0.32 $ & $   2.08 \pm   0.01     $ & $   4.24 \pm   0.14 $ & $   7.78 \pm   0.14 $  \\
8853-57459-322  &  157.7832  &   35.8713  &    0.64  & $   0.21 \pm   0.13 $ &        4.83  &      0.61   &  170.93  & $     116.14 \pm   4.24  $ & $ 4.97 \pm     0.25     $ & $ 34.26 \pm 7.08 $ & $     8.81 \pm 0.09 $ & $   1.40 \pm   0.04     $ & $   4.27 \pm   0.07 $ & $   7.64 \pm   0.11 $  \\
8179-57429-876  &  161.4052  &   56.7378  &    0.64  & $   0.00 \pm   0.08 $ &        5.58  &      0.66   &  141.42  & $     149.42 \pm   3.52  $ & $ 4.19 \pm     0.16     $ & $ 56.69 \pm 9.47 $ & $     8.88 \pm 0.03 $ & $   1.52 \pm   0.02     $ & $   4.16 \pm   0.05 $ & $   7.91 \pm   0.07 $  \\
6444-56339-270  &  162.2911  &   33.1740  &    0.77  & $   0.00 \pm   0.14 $ &        5.61  &      0.64   &   96.97  & $      48.57 \pm   1.79  $ & $ 5.55 \pm     0.28     $ & $ 23.57 \pm 5.80 $ & $     9.02 \pm 0.28 $ & $   1.22 \pm   0.04     $ & $   4.37 \pm   0.14 $ & $   7.50 \pm   0.17 $  \\
7396-56809-134  &  171.9186  &   45.2847  &    0.68  & $   0.10 \pm   0.14 $ &        6.22  &      0.85   &   61.90  & $      75.28 \pm   1.81  $ & $ 3.25 \pm     0.16     $ & $ 24.33 \pm 4.97 $ & $     8.76 \pm 0.45 $ & $   1.28 \pm   0.02     $ & $   4.36 \pm   0.07 $ & $   7.35 \pm   0.09 $  \\
7399-57162-661  &  172.3675  &   49.3908  &    0.85  & $   0.00 \pm   0.21 $ &        3.84  &      0.98   &   69.92  & $      34.42 \pm   2.89  $ & $ 4.04 \pm     0.44     $ & $ 48.84 \pm 36.28 $ & $     9.04 \pm 0.11 $ & $   1.18 \pm   0.08     $ & $   4.19 \pm   0.20 $ & $   7.83 \pm   0.16 $  \\
7400-57134-924  &  175.2757  &   50.4547  &    0.68  & $   0.11 \pm   0.11 $ &        3.29  &      0.83   &   72.34  & $     140.85 \pm   4.90  $ & $ 4.42 \pm     0.22     $ & $ 99.29 \pm 34.21 $ & $     9.42 \pm 0.44 $ & $   1.56 \pm   0.03     $ & $   4.06 \pm   0.09 $ & $   8.22 \pm   0.12 $  \\
8230-57430-196  &  176.7698  &   54.5643  &    0.73  & $   0.75 \pm   0.18 $ &        3.32  &      0.75   &  143.37  & $     1220.27 \pm   3.35  $ & $ 4.81 \pm     0.02     $ & $ 40.63 \pm 0.50 $ & $     9.01 \pm 0.19 $ & $   2.57 \pm   0.03     $ & $   4.22 \pm   0.08 $ & $   7.93 \pm   0.14 $  \\
8228-57423-302  &  178.8998  &   52.9714  &    0.68  & $   0.36 \pm   0.14 $ &        5.20  &      0.99   &   34.67  & $     257.86 \pm   5.57  $ & $ 3.89 \pm     0.12     $ & $ 29.39 \pm 7.41 $ & $     9.37 \pm 0.14 $ & $   1.82 \pm   0.02     $ & $   4.30 \pm   0.16 $ & $   7.65 \pm   0.20 $  \\
5979-56329-352  &  185.5227  &   23.3482  &    0.68  & $   0.00 \pm   0.17 $ &        3.48  &      0.74   &   70.33  & $      39.23 \pm   1.20  $ & $ 4.93 \pm     0.23     $ & $ 47.65 \pm 10.25 $ & $     8.81 \pm 0.06 $ & $   1.00 \pm   0.03     $ & $   4.19 \pm   0.09 $ & $   7.86 \pm   0.12 $  \\
6667-56412-226  &  188.5501  &   46.6314  &    0.72  & $   0.24 \pm   0.20 $ &        4.27  &      0.88   &   89.36  & $     103.67 \pm   2.83  $ & $ 5.18 \pm     0.19     $ & $ 59.93 \pm 17.06 $ & $     8.53 \pm 0.15 $ & $   1.49 \pm   0.03     $ & $   4.15 \pm   0.19 $ & $   8.01 \pm   0.19 $  \\
7415-57097-658  &  190.2230  &   49.4394  &    0.69  & $   0.00 \pm   0.11 $ &        4.20  &      0.59   &  191.21  & $      59.57 \pm   1.52  $ & $ 5.32 \pm     0.19     $ & $ 51.96 \pm 6.53 $ & $     8.18 \pm 0.04 $ & $   1.20 \pm   0.03     $ & $   4.17 \pm   0.05 $ & $   7.88 \pm   0.09 $  \\
7417-56753-685  &  192.6378  &   46.3512  &    0.60  & $   0.00 \pm   0.09 $ &       10.72  &      0.52   &  124.31  & $      92.56 \pm   3.73  $ & $ 5.80 \pm     0.33     $ & $ 67.83 \pm 13.19 $ & $     8.59 \pm 0.16 $ & $   1.24 \pm   0.04     $ & $   4.12 \pm   0.03 $ & $   8.02 \pm   0.05 $  \\
8222-57431-673  &  193.1308  &   52.6212  &    0.62  & $   0.15 \pm   0.13 $ &        5.66  &      0.70   &  189.42  & $     130.15 \pm   3.94  $ & $ 4.85 \pm     0.19     $ & $ 49.73 \pm 7.33 $ & $     8.59 \pm 0.48 $ & $   1.43 \pm   0.03     $ & $   4.18 \pm   0.07 $ & $   7.84 \pm   0.10 $  \\
8223-57429-50   &  198.9668  &   50.6363  &    0.64  & $   0.00 \pm   0.17 $ &        5.10  &      0.73   &  157.57  & $      52.54 \pm   2.94  $ & $ 5.47 \pm     0.41     $ & $ 36.57 \pm 6.15 $ & $     9.05 \pm 0.09 $ & $   1.06 \pm   0.06     $ & $   4.25 \pm   0.07 $ & $   7.71 \pm   0.09 $  \\
8211-57423-304  &  203.3647  &   50.2998  &    0.62  & $   0.06 \pm   0.10 $ &        5.07  &      0.87   &   73.74  & $     110.68 \pm   2.58  $ & $ 3.74 \pm     0.13     $ & $ 56.54 \pm 10.75 $ & $     9.33 \pm 0.07 $ & $   1.36 \pm   0.02     $ & $   4.16 \pm   0.07 $ & $   7.86 \pm   0.10 $  \\
7428-56781-95   &  209.5281  &   46.1953  &    0.62  & $   0.21 \pm   0.10 $ &        6.11  &      0.82   &   99.58  & $     222.14 \pm   6.04  $ & $ 4.39 \pm     0.20     $ & $ 71.50 \pm 13.58 $ & $     9.00 \pm 0.47 $ & $   1.65 \pm   0.03     $ & $   4.11 \pm   0.06 $ & $   8.05 \pm   0.09 $  \\
6932-56397-238  &  214.8633  &   53.1171  &    0.60  & $   0.67 \pm   0.19 $ &        3.18  &      0.96   &   44.84  & $     329.07 \pm   3.65  $ & $ 3.55 \pm     0.05     $ & $ 28.22 \pm 2.14 $ & $     8.79 \pm 0.20 $ & $   1.80 \pm   0.01     $ & $   4.31 \pm   0.15 $ & $   7.67 \pm   0.19 $  \\
8498-57105-124  &  222.6245  &   39.2227  &    0.61  & $   0.00 \pm   0.11 $ &        4.64  &      0.83   &   75.76  & $      84.41 \pm   2.04  $ & $ 4.46 \pm     0.16     $ & $ 85.54 \pm 20.86 $ & $     9.18 \pm 0.27 $ & $   1.22 \pm   0.02     $ & $   4.09 \pm   0.09 $ & $   8.09 \pm   0.13 $  \\
3877-55365-528  &  224.6807  &   28.8287  &    0.69  & $   0.04 \pm   0.16 $ &        4.23  &      0.94   &   41.25  & $      46.80 \pm   1.30  $ & $ 5.75 \pm     0.22     $ & $ 41.35 \pm 9.22 $ & $     8.47 \pm 0.47 $ & $   1.09 \pm   0.03     $ & $   4.22 \pm   0.13 $ & $   7.83 \pm   0.16 $  \\
5143-55828-182  &  322.4144  &    0.8864  &    0.72  & $   0.00 \pm   0.17 $ &        3.24  &      0.74   &  103.87  & $      21.59 \pm   1.17  $ & $ 6.55 \pm     0.48     $ & $ 58.13 \pm 20.80 $ & $     8.22 \pm 0.45 $ & $   0.81 \pm   0.05     $ & $   4.15 \pm   0.12 $ & $   7.99 \pm   0.11 $  \\
\enddata
\end{deluxetable*}
\end{longrotatetable}

\section{Results}
\label{sec:results}

\subsection{Star-forming Main Sequence}

\begin{figure}
\epsscale{1.0}
\plotone{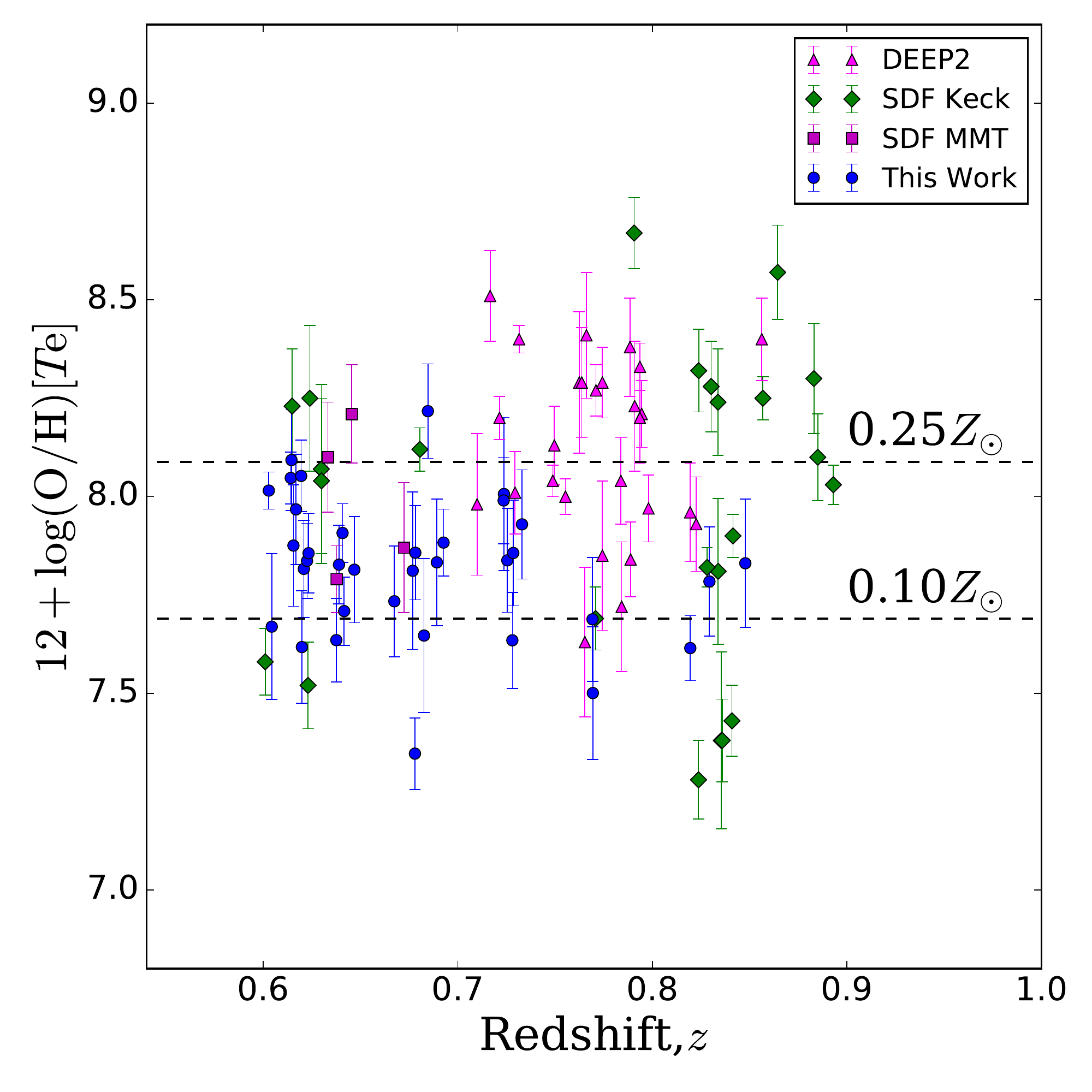}
\caption{The distributions of $T_{\rm e}$-based metallicities and redshifts for some samples of $\oiii\lambda4363$-detected SFGs at $0.6<z<0.9$, including 35 SFGs in our eBOSS sample (the blue circles), 28 SFGs from the DEEP2 sample (the magenta triangles) \citep{Ly2015a}, 4 SDF MMT/Hectospec sample (the purple squares), and 24 SDF Keck/DEIMOS sample (the green diamonds) \citep{Ly2016}.
\label{fig:redshift_Z}}
\end{figure}

In Figure \ref{fig:redshift_Z}, we show the distributions of $T_{\rm e}$-based metallicities versus redshifts for some samples of $\oiii\lambda4363$-detected SFGs at $0.6<z<0.9$, including 35 SFGs in our eBOSS sample (the blue circles), 28 SFGs from the DEEP2 sample (the magenta triangles) from \cite{Ly2015a}, 4 SDF MMT/Hectospec sample (the purple squares), and 24 SDF Keck/DEIMOS sample (the green diamonds) from \cite{Ly2016}. The redshifts of our SFGs span from 0.6 to 0.9. Compared with other samples, most of our SFGs (24/35) are located in $0.6 \leq z \leq 0.7$, while most of the DEEP2 and SDF sample range in $0.7 - 0.9$.

\begin{figure}
\epsscale{1.0}
\plotone{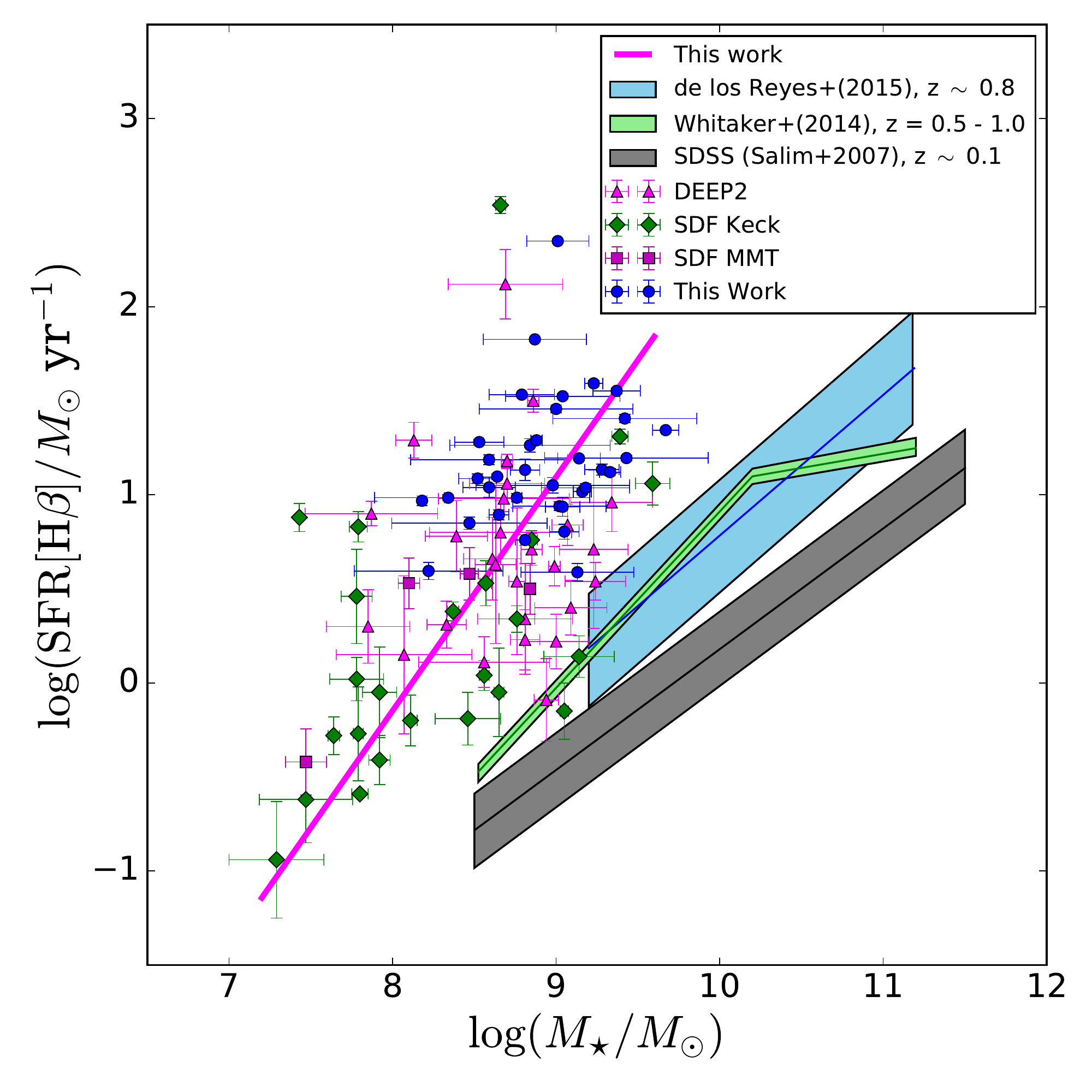}
\caption{The star-forming main sequence relation for some samples of $\oiii\lambda4363$-detected SFGs at $0.6<z<0.9$, which symbols are same as Figure \ref{fig:redshift_Z}. The solid magenta line is the best linear fitting relation for above metal-poor galaxies.  Overlaid as blue and green stripes are the results based on $\ha$-selected galaxies at $z\sim 0.8$ \citep{Reyes2015} and mass-selected SFGs at $z=0.5-1.0$ \citep{Whitaker2014}, respectively. The local star-forming main sequence derived from the SDSS sample of SFGs is also shown in the grey stripe \citep{Salim2007}.
\label{fig:Mass_SFR}}
\end{figure}

In Figure \ref{fig:Mass_SFR}, we plot the metallicity-corrected SFRs derived with $\hb$ luminosities as a function of the stellar masses from SED fitting to compare against other metal-poor SFG samples. The symbols are the same as in Figure \ref{fig:redshift_Z}.  We also perform the linear fitting for these metal-poor galaxies, and give the best relation as,
\begin{equation}
\label{eq:mass_sfr}
\begin{split}
    {\rm log(SFR}/M_\odot {\rm yr^{-1}}) = 1.24({\pm 0.11}) \ {\rm log}(M_\star/M_\odot) \\
     - 10.09({\pm 0.10}),
\end{split}
\end{equation}
shown with solid magenta line.  Overlaid as blue and green stripes are the results based on $\ha$-selected galaxies at $z\sim 0.8$ \citep{Reyes2015} and mass-selected SFGs at $z=0.5-1.0$ \citep{Whitaker2014}, respectively. The local star-forming main sequence derived from the SDSS sample of SFGs is also shown in the grey stripe \citep{Salim2007}. The difference between the main sequence relations in local and high redshift universe is also consistent with the result that the rate of declining SFR with redshift is not a strong function of stellar mass \citep{Zheng2007}.  Compared with the SFGs with normal metallicities, the metal-poor SFGs in our sample are found to have more intense star formation activities. If extrapolating the $M_\star$ -- SFR relation to lower $M_\star$, we find the SFRs in our sample are enhanced by 1.0 -- 3.0 dex above the $M_\star$ -- SFR relation in local universe \citep{Salim2007}, and are 0.5 -- 2.0 dex higher than the nomarl SFGs at $z=0.5-1.0$ \citep{Whitaker2014}  or  $z \sim 0.8$ \citep{Reyes2015}. The coverage of stellar masses and SFRs of our eBOSS SFGs is similar to the DEEP2 sample.

\subsection{Stellar Mass -- Metallicity Relation}

\begin{figure}
\epsscale{1.0}
\plotone{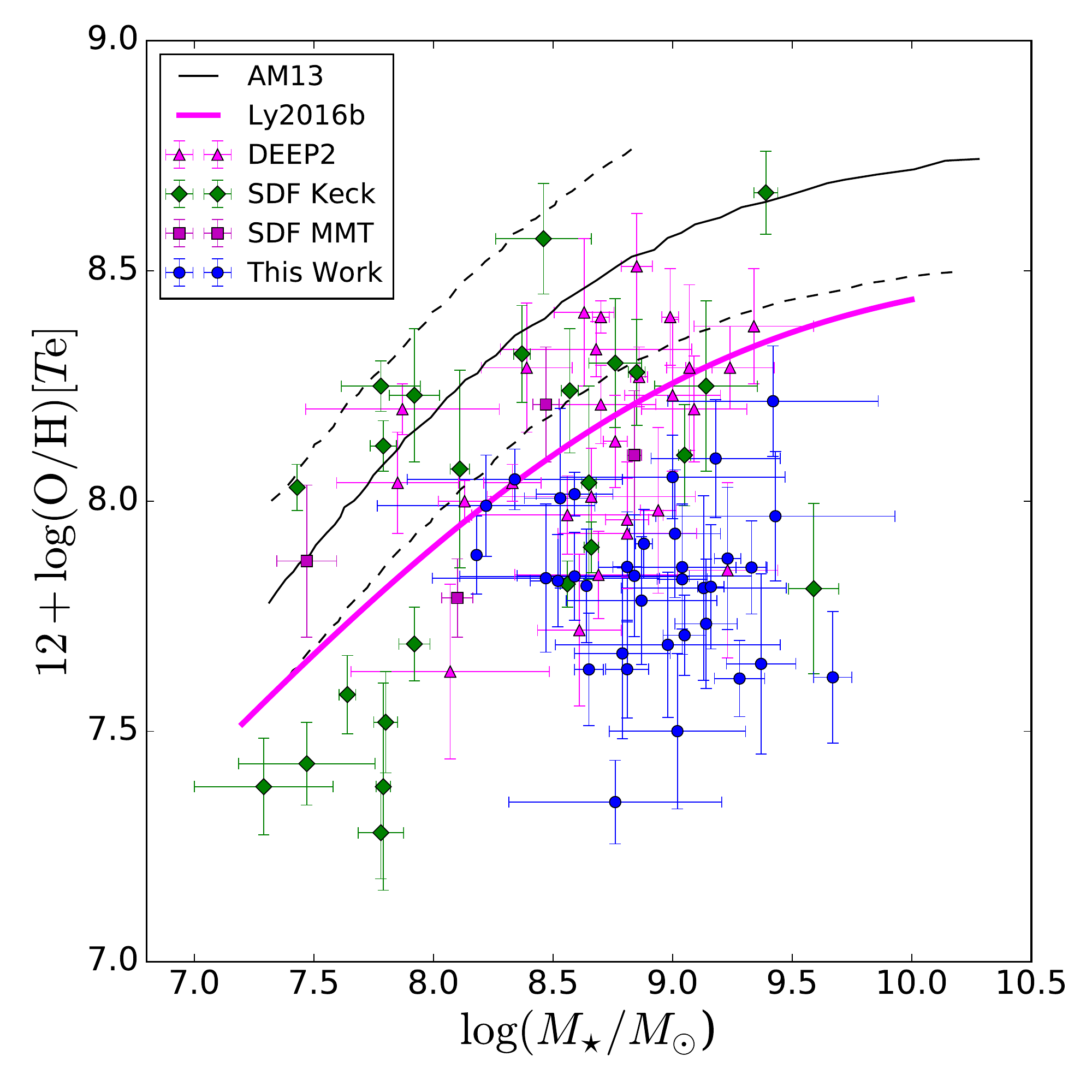}
\caption{The relation between gas-phase metallicity and stellar mass (i.e., MZR) for some samples of $\oiii\lambda4363$-detected SFGs at $0.6<z<0.9$, which symbols are same as in Figure \ref{fig:redshift_Z}. For a direct comparison, the local MZR and its limits derived from the stacked spectra by \citet{Andrews2013} are given as black solid and dashed lines, while the MZR for SFGs at $0.5<z<1.0$ derived by \cite{Ly2016a} is shown as the solid magenta line.
\label{fig:Mass_Z}}
\end{figure}

In Figure \ref{fig:Mass_Z}, we plot the relation between gas-phase metallicity and stellar mass (i.e., MZR) for some samples of $\oiii\lambda4363$-detected SFGs at $0.6<z<0.9$, including metal-poor SFGs in our eBOSS sample and three other sample, which symbols are same as in Figure \ref{fig:redshift_Z} and \ref{fig:Mass_SFR}. The black solid and dashed lines represent the local MZR and its limits derived from the stacked spectra by \cite{Andrews2013} based on $T_{\rm e}$ metallicity calculation, respectively. The solid magenta line represents the MZR for SFGs at $0.5<z<1.0$ derived by \cite{Ly2016a}. As shown in the figure, most of our eBOSS metal-poor SFGs show lower metallicities than the MZRs in \cite{Andrews2013} and \cite{Ly2016a} systematically, by about 0.7 dex and 0.3 dex, respectively.  Besides, we note that the metallicities in eBOSS SFGs are lower than the metallicities in the DEEP2 sample about 0.37 dex.

In the previous studies, \cite{Savaglio2005} identified a strong correlation between stellar mass and metallicity for 56 galaxies based on R23 method at the high-$z$ universe, and for the first time, found clear evidence for the cosmic evolution of MZR. \cite{Lian2016} also used the R23 method to compare the metallicities of BCDs in local and intermediate-$z$ ($0.2 < z < 0.5$) universe, but did not find a significant deviation in MZR.  Determining the metallicities for 66 $\oiii\lambda4363$-detected galaxies with the electron temperature metallicity calibration, \cite{Ly2016a} found the MZR at high redshift have lower metallicities, about 0.25 dex, than the MZR in \cite{Andrews2013} at all stellar masses. In this work, our SFGs also have systematical lower metallicities than the local MZR, confirming the existence of the cosmic evolution of MZR with redshift.  Meanwhile, our SFGs also locate below the MZR at $0.5<z<1.0$ in \cite{Ly2016a}, which may be caused by the higher SFRs, higher stellar masses and/or the uncertainties of stellar mass measurements. Because of the shallow limited-magnitude in eBOSS, these relatively more massive ($8.0 \le {\rm log}(M_\star/M_\odot) \le 9.5$) metal-poor galaxies may be preferably chosen in the sample selection \citep{Izotov2014}.

\subsection{Fundamental Metallicity Relation}
\label{subsec:fmr}

\begin{figure}
\epsscale{1.0}
\plotone{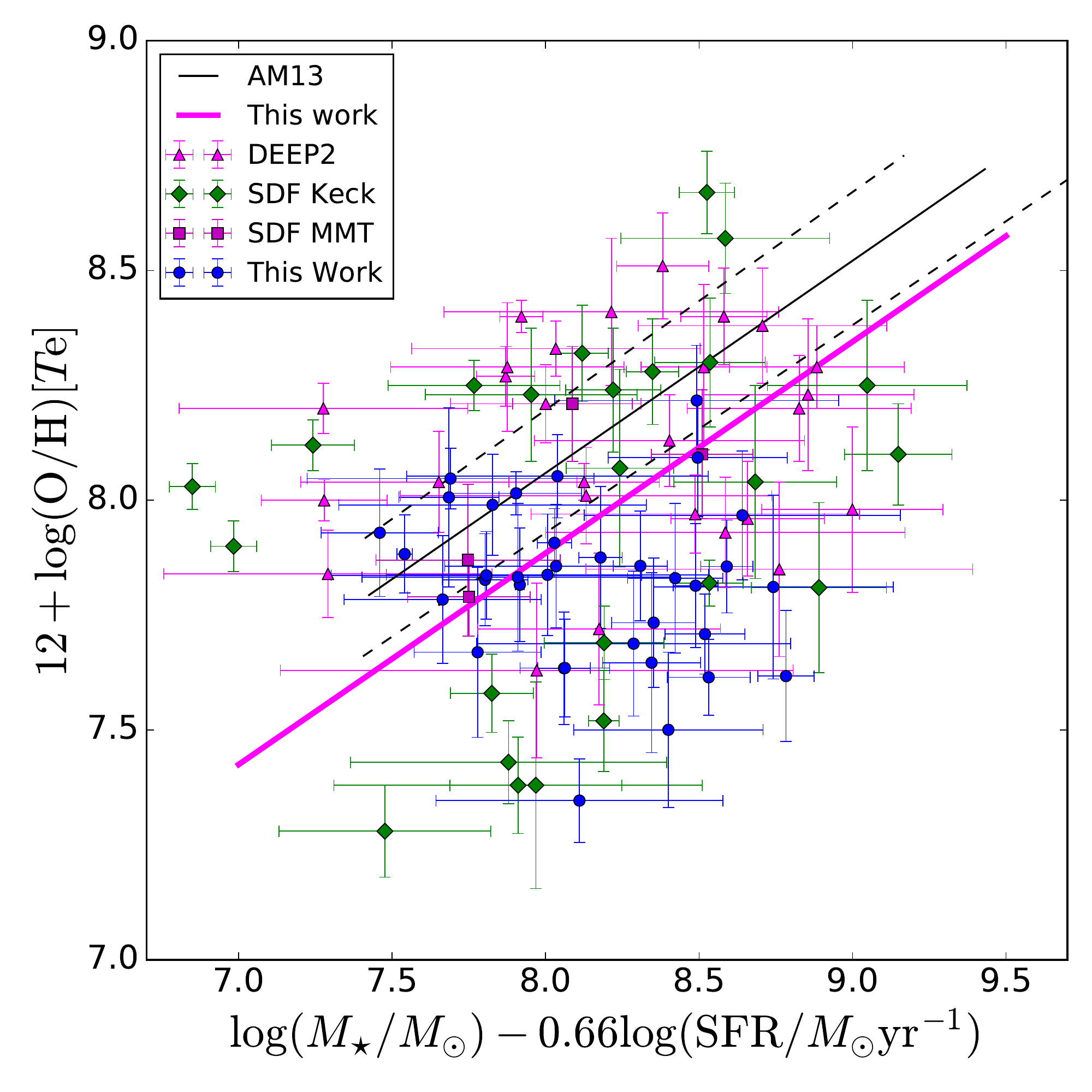}
\caption{The fundamental metallicity relation (FMR) for some samples of $\oiii\lambda4363$-detected SFGs at $0.6<z<0.9$, which symbols are same as in Figure \ref{fig:redshift_Z}. For a direct comparison, the local FMR derived from the stacked spectra by \cite{Andrews2013} is also given as the black solid and dashed lines. The best FMR relation we fit with the same slope of 0.43 in \cite{Andrews2013} is shown as the solid magenta line, which has an offset of 0.16 dex below the local FMR.
\label{fig:FMR}}
\end{figure}

The stellar mass -- SFR -- metallicity relation (FMR) is constructed by \cite{Mannucci2010} and \cite{Andrews2013} to significantly decrease the scatter in MZR, which indicates that the metallicity to be anti-correlated with SFR at a fixed stellar mass. In Figure \ref{fig:FMR}, we show the FMR for some samples of \oiii$\lambda4363$-detected SFGs at $0.6<z<0.9$, which symbols are same as in Figure \ref{fig:redshift_Z}, \ref{fig:Mass_SFR} and \ref{fig:Mass_Z}.  For a direct comparison, we adopt the SFR coefficient $\alpha = 0.66$ from \cite{Andrews2013}, and give the local FMR with its limits as the black solid and dashed lines.  We can find that 10/35 of our metal-poor SFGs can be well fitted by the local FMR within a deviation of 0.14 dex, while the other 25/35 SFGs have the metallicities below the local FMR with an average value 0.43 dex. Furthermore, we also use the linear relation with the same slope of 0.43 in \cite{Andrews2013} to fit all these data, and find the best relation (shown as the solid magenta line) has an offset of about 0.16 dex below the local FMR, with a dispersion (standard deviation of the residuals) about 0.32 dex.

Our results confirm the existence of FMR, which means that the metal-poor galaxies with higher SFRs have lower metallicities at a same stellar mass. The offset of FMR also indicates the trend that, similar to MZR cosmic evolution, FMR evolves toward lower metallicity at a fixed stellar mass and SFR in the earlier universe. We note that our SFGs have smaller $D_{\rm n}4000$ indices ($0.5 \le D_{\rm n}4000 \le 1.0$) with a median value of 0.8. The cosmic evolution of FMR may be caused by the fact that SFGs at intermediate-$z$ and high-$z$ have much younger stellar populations than those in the local universe.  This result is consistent with the result in \cite{Lian2015} that galaxies with smaller $D_{\rm n}4000$ typically have lower metallicity at a fixed stellar mass, suggesting that the galaxy stellar age plays an essential role in the MZR and FMR for high-redshift SFGs.

\subsection{Discussion for Metallicity and Stellar Mass Determinations}
\label{subsec:discussion_z_mass}

In the process to determine the metallicity with electron temperature, we usually need the $\sii\lambda6717/\sii\lambda6731$ ratio to estimate the electron density $n_e$, and then use the $\oiii\lambda\lambda4959,5007/\oiii\lambda4363$ and $n_e$ to determine the $T_{\rm e}(\oiii)$.  However, we note that the $n_e$ effect on metallicity becomes obvious only at $n_e > 10^4 \, \rm cm^{-3}$. If we assume the $n_e = 1,000\, \rm cm^{-3}$, the metallicities of the SFGs in our sample have a slight offset, about 0.02 dex above those with $n_e = 100\, \rm cm^{-3}$.  As to the $T_{\rm e}(\oiii)$ determination, the $T_{\rm e}(\oii)$ can be derived with $\oii\lambda3727/\oii\lambda\lambda7320,7330$ ratio. Due to the absence of $\oii\lambda\lambda7320,7330$ in our spectra,  we follow the method in \cite{Nicholls2014} to determine the $T_{\rm e}(\oii)$.  If calculating the metallicities using the $T_{\rm e}(\oii)$ -- $T_{\rm e}(\oiii)$ relation from \cite{Izotov2012} and \cite{Andrews2013},  we will yield higher $T_{\rm e}(\oii)$ values with an average offset about 0.03 dex, which will result in lower metallicities about 0.05 dex. 

\begin{figure}
\epsscale{1.0}
\plotone{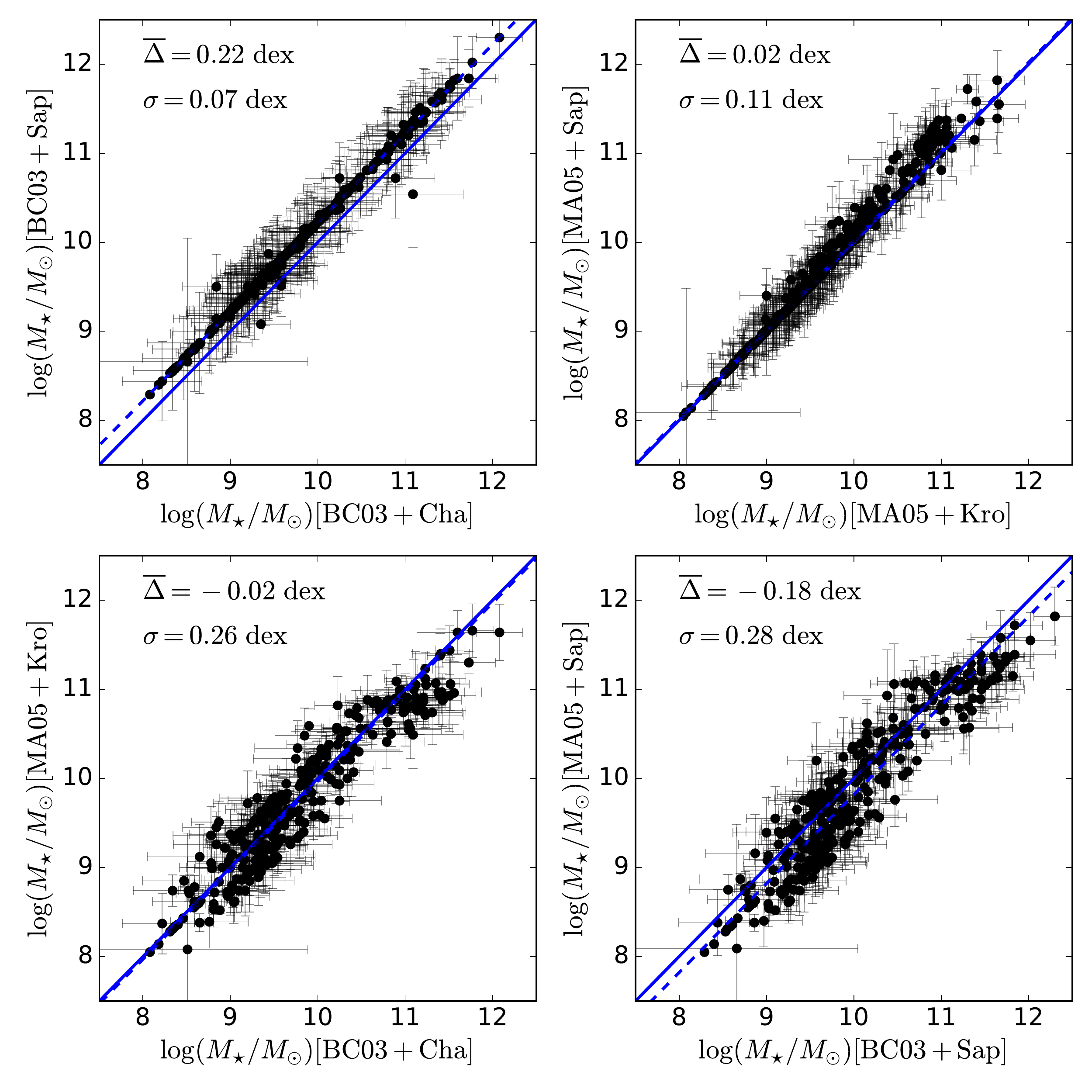}
\caption{The comparison of stellar mass determination with different SSP libraries and IMFs for the subsample of 319 galaxies selected in Section \ref{sub:sample}. The \cite{Bruzual2003} and \cite{Maraston2005} libraries are shown as BC03 and MA05, the \cite{Chabrier2003}, \cite{Kroupa2001}  and \cite{Salpeter1955} IMFs are abbreviated as Cha, Kro and Sap, repectively. The blue solid lines indicate equality between the stellar masses derived with different parameters. The median values ($\overline\Delta$, offset, also shown as blue dashed lines) and the standard deviation ($\sigma$) of residuals are also shown in the legends.
\label{fig:comp_mass}}
\end{figure}

In order to make a direct comparison with the metal-poor SFGs selected from SDF and DEEP2 surveys, we assume the \cite{Chabrier2003} IMF and \cite{Bruzual2003} SSPs, just following \cite{Ly2014} and \cite{Ly2016}, to estimate the stellar masses. However, different assumptions on IMFs and SSPs will lead to some difference in the stellar mass determination \citep[e.g.,][]{Bruzual2003,Brinchmann2008,Parikh2018}.  In Figure \ref{fig:comp_mass}, we show the comparison of stellar mass determination with two different SSP libraries \citep{Bruzual2003,Maraston2005} and three IMFs \citep{Salpeter1955,Kroupa2001,Chabrier2003} for the subsample of 319 galaxies selected in Section \ref{sub:sample}. If using the \cite{Bruzual2003} SSPs, we note that the stellar masses estimated by \cite{Salpeter1955} IMF have a systematical offset about 0.22 dex above these by \cite{Chabrier2003} IMF, consistent with \cite{Sanchez2016a}, while have a smallest dispersion about 0.07 dex.  If we adopt the \cite{Salpeter1955} IMF, the offset between stellar masses with \cite{Bruzual2003} and \cite{Maraston2005} SSPs is about -0.18 dex, and the dispersion is about 0.28 dex. However, as \cite{Bruzual2003} pointed out, the \cite{Chabrier2003} IMF is physically motivated, and it shows a better fitting to the counts of low-mass stars and brown dwarfs in the Galactic disc. Compared with the \cite{Chabrier2003} IMF, \cite{Salpeter1955} IMF leads to a systematical overestimation of stellar mass. In brief, the different assumptions in metallicity calibration and mass determination will have a nearly nonsignificant effect on the MZR and FMR for our metal-poor SFGs in eBOSS.

\section{Summary}
\label{sec:summary}

In this work, we use the spectroscopic data of eBOSS in SDSS DR14 dataset to search the metal-poor galaxies with $\oiii\lambda4363$ detection at $0.6 < z < 0.9$, and to explore the cosmic evolution of MZR and FMR. We have determined the metallicity with electron temperature method based on $\oiii\lambda4363/\oiii\lambda5007$ ratio. The stellar masses are derived from the SED fitting with multi-photometric measurements, and the SFRs are estimated from the $\hb$ luminosities. The primary results are summarized as follows.

\begin{itemize}
\item We select a sample of 35 metal-poor star-forming galaxy candidates with $\rm S/N(\oiii\lambda4363) \ge 3.0$ and has a median value of $\oh = 7.83$ in range from 7.35 to 8.22 (Figure \ref{fig:redshift_Z}).  Nine SFGs are found to be extremely metal-poor with $\oh \le 7.69$, in which the most metal-deficient galaxy is 7.35 $\pm$ 0.09, about 1/20 $Z_\odot$.

\item We find the SFRs of our SFGs are enhanced by 1.0 -- 3.0 dex above the $M_\star$ -- SFR relation \citep{Salim2007} in local universe (Figure \ref{fig:Mass_SFR}). Based on a few of metal-poor galaxy samples, we create a new $M_\star$ -- SFR relation for these galaxies at $0.6 \le z \le 0.9$, shown as Equation \ref{eq:mass_sfr}.

\item Our SFGs have systematical lower metallicities than the local MZR in \cite{Andrews2013}, confirming the existence of the cosmic evolution of MZR with redshift reported in previous studies (Figure \ref{fig:Mass_Z}). Furthermore, we also find the metallicities of these metal-poor SFGs have a systematical offset about 0.16 dex below the local FMR (Figure \ref{fig:FMR}), indicating that galaxies have lower metallicites at a fixed stellar mass and SFR in the earlier universe.
\end{itemize}

We attribute the cosmic evolution of FMR to the stellar age of galaxies at different redshifts (Section \ref{subsec:fmr}). In our next work, we will stack the spectra, for all ELGs in the eBOSS survey, in bins of stellar mass, SFR and $D_{\rm n}4000$, to significantly enhance the S/N of emission lines, and then mainly explore the cosmic evolution of MZR and FMR.

\acknowledgments
We thank the referee for his/her thorough reading and valuable suggestions. This work is supported by the National Natural Science Foundation of China (NSFC, Nos. 11873032, 11173016, 11320101002, 11421303, and 11433005), the Research Fund for the Doctoral Program of Higher Education of China (Nos. 20133207110006), and the National Key R\&D Program of China (2015CB857004, 2016YFA0400702, 2017YFA0402600).  

Funding for the Sloan Digital Sky Survey IV has been provided by the Alfred P. Sloan Foundation, the U.S. Department of Energy Office of Science, and the Participating Institutions. SDSS acknowledges support and resources from the Center for High-Performance Computing at the University of Utah. The SDSS web site is www.sdss.org.

The Legacy Surveys consist of three individual and complementary projects: the Dark Energy Camera Legacy Survey (DECaLS; NOAO Proposal ID \# 2014B-0404; PIs: David Schlegel and Arjun Dey), the Beijing-Arizona Sky Survey (BASS; NOAO Proposal ID \# 2015A-0801; PIs: Zhou Xu and Xiaohui Fan), and the Mayall z-band Legacy Survey (MzLS; NOAO Proposal ID \# 2016A-0453; PI: Arjun Dey). DECaLS, BASS and MzLS together include data obtained, respectively, at the Blanco telescope, Cerro Tololo Inter-American Observatory, National Optical Astronomy Observatory (NOAO); the Bok telescope, Steward Observatory, University of Arizona; and the Mayall telescope, Kitt Peak National Observatory, NOAO. The Legacy Surveys project is honored to be permitted to conduct astronomical research on Iolkam Du'ag (Kitt Peak),a mountain with particular significance to the Tohono O'odham Nation.

NOAO is operated by the Association of Universities for Research in Astronomy (AURA) under a cooperative agreement with the National Science Foundation.

This project used data obtained with the Dark Energy Camera (DECam), which was constructed by the Dark Energy Survey (DES) collaboration. Funding for the DES Projects has been provided by the U.S. Department of Energy, the U.S. National Science Foundation, the Ministry of Science and Education of Spain, the Science and Technology Facilities Council of the United Kingdom, the Higher Education Funding Council for England, the National Center for Supercomputing Applications at the University of Illinois at Urbana-Champaign, the Kavli Institute of Cosmological Physics at the University of Chicago, Center for Cosmology and Astro-Particle Physics at the Ohio State University, the Mitchell Institute for Fundamental Physics and Astronomy at Texas A\&M University, Financiadora de Estudos e Projetos, Fundacao Carlos Chagas Filho de Amparo, Financiadora de Estudos e Projetos, Fundacao Carlos Chagas Filho de Amparo a Pesquisa do Estado do Rio de Janeiro, Conselho Nacional de Desenvolvimento Cientifico e Tecnologico and the Ministerio da Ciencia, Tecnologia e Inovacao, the Deutsche Forschungsgemeinschaft and the Collaborating Institutions in the Dark Energy Survey. The Collaborating Institutions are Argonne National Laboratory, the University of California at Santa Cruz, the University of Cambridge, Centro de Investigaciones Energeticas, Medioambientales y Tecnologicas-Madrid, the University of Chicago, University College London, the DES-Brazil Consortium, the University of Edinburgh, the Eidgenossische Technische Hochschule (ETH) Zurich, Fermi National Accelerator Laboratory, the University of Illinois at Urbana-Champaign, the Institut de Ciencies de l'Espai (IEEC/CSIC), the Institut de Fisica d'Altes Energies, Lawrence Berkeley National Laboratory, the Ludwig-Maximilians Universitat Munchen and the associated Excellence Cluster Universe, the University of Michigan, the National Optical Astronomy Observatory, the University of Nottingham, the Ohio State University, the University of Pennsylvania, the University of Portsmouth, SLAC National Accelerator Laboratory, Stanford University, the University of Sussex, and Texas A\&M University.

BASS is a key project of the Telescope Access Program (TAP), which has been funded by the National Astronomical Observatories of China, the Chinese Academy of Sciences (the Strategic Priority Research Program "The Emergence of Cosmological Structures" Grant \# XDB09000000), and the Special Fund for Astronomy from the Ministry of Finance. The BASS is also supported by the External Cooperation Program of Chinese Academy of Sciences (Grant \# 114A11KYSB20160057), and Chinese National Natural Science Foundation (Grant \# 11433005).

The Legacy Survey team makes use of data products from the Near-Earth Object Wide-field Infrared Survey Explorer (NEOWISE), which is a project of the Jet Propulsion Laboratory/California Institute of Technology. NEOWISE is funded by the National Aeronautics and Space Administration.

The Legacy Surveys imaging of the DESI footprint is supported by the Director, Office of Science, Office of High Energy Physics of the U.S. Department of Energy under Contract No. DE-AC02-05CH1123, by the National Energy Research Scientific Computing Center, a DOE Office of Science User Facility under the same contract; and by the U.S. National Science Foundation, Division of Astronomical Sciences under Contract No. AST-0950945 to NOAO.

\bibliography{eboss_ms}

\end{document}